\documentclass[aps,preprint]{revtex4}
\usepackage[dvips]{graphicx}
\usepackage{amssymb}
\begin{document}
\draft

\title{\bf Are scattering properties of graphs uniquely connected to
    their shapes?
}
\author{Oleh Hul,$^{1}$ Micha{\l} {\L}awniczak,$^{1}$ Szymon Bauch,$^{1}$ \\ Adam Sawicki,$^{2,3}$ Marek Ku\'s,$^{2}$
and Leszek Sirko$^{1}$}

\address{$^{1}$Institute of Physics, Polish Academy of Sciences, Aleja  Lotnik\'{o}w 32/46, 02-668 Warszawa, Poland\\
$^{2}$Center for Theoretical Physics, Polish Academy of Sciences, Aleja  Lotnik\'{o}w 32/46, 02-668 Warszawa, Poland\\
$^{3}$School of Mathematics, University of Bristol, University
Walk, Bristol BS8 1TW, UK}

\date{May 23, 2012}
\bigskip

\begin{abstract}
The famous question of Mark Kac "Can one hear the shape of a drum?" addressing
the unique connection between the shape of a planar region and the spectrum of
the corresponding Laplace operator can be legitimately extended to scattering
systems. In the modified version one asks whether the geometry of a vibrating
system can be determined by scattering experiments. We present the first
experimental approach to this problem in the case of microwave graphs
(networks) simulating quantum graphs. Our experimental results strongly
indicate a negative answer. To demonstrate this we consider scattering from a
pair of isospectral microwave networks consisting of vertices connected by
microwave coaxial cables and extended to scattering systems by connecting leads
to infinity to form \textit{isoscattering} networks. We show that the
amplitudes and phases of the determinants of the scattering matrices of such
networks are the same within the experimental uncertainties. Furthermore, we
demonstrate that the scattering matrices of the networks are conjugated by the,
so called, transplantation relation.

\end{abstract}


\bigskip
\maketitle

The problem of isospectrality goes back to 1966 when Marc Kac posed a famous
question "Can one hear the shape of a drum?" \cite{Kac66}. It addressed the
issue of uniqueness of the spectrum of the Laplace  on the planar domain with
Dirichlet boundary conditions. The answer was not found until 1992 when Gordon,
Webb, and Wolpert \cite{Gordon92a,Gordon92b} using the Sunada's theorem
\cite{Sunada85} found a way to construct pairs of isospectral domains in
$\mathbb{R}^{2}$. An experimental confirmation that the shape of a drum can not
be heard was presented by Sridhar and Kudrolli \cite{Sridhar1994} for a pair of
isospectral microwave cavities.

Inability of determining the shape from the spectrum alone does not preclude
possibilities of distinguishing one drum from another in more sophisticated
experiments. Indeed, basing on numerical simulations Okada et al.\
\cite{OSTH05} conjectured that isospectral domains constructed by Gordon, Webb
and Wolpert can be in fact discriminated in scattering experiments looking at
poles of the scattering matrices.

Original question of Mark Kac can be posed for other vibrating systems.
Gutkin and Smilansky \cite{Gutkin01} considered the problem of isospectrality
in the context of quantum graphs. They proved that one can recover a graph
from its spectrum if the lengths of its bonds are incommensurate. Their
result gives a room for existence of graphs with different metric and
topological properties but the same spectrum. Up to now there is only one
method of construction of isospectral graphs \cite{Band09,PB09} where the
authors extended the well known Sunada's approach. The method is based on the
elements of representation theory and its direct corollary ensures the
existence of transplantation between isospectral graphs. Roughly speaking in
the process of transplantation one graph is divided into smaller building
blocks which are then reassembled to form the second one of a different
shape. The method provides also correct boundary conditions at vertices of
the new graph. As a result to every eigenfunction on the first graph an
eigenfunction with the same eigenvalue on the second one is assigned. The
procedure is reminiscent of the one used in designing isospectral planar
domains where the 'drum' is cut into subdomains which are then rearranged
into a new one with the same spectrum. Following the conjecture of Okada's et
al.\ one can thus ask whether the geometry of a quantum graph can be
determined by scattering experiments.

The negative answer was given by Band, Sawicki and Smilansky
\cite{Sawicki10,Sawicki10b}. They extended the theory of isospectrality to
scattering systems by considering isospectral quantum graphs with attached
infinite leads and developed a method of constructing \textit{isoscattering}
pairs of graphs for which scattering matrices have the same polar structure. In
particular, they showed that any pair of isospectral quantum graphs obtained by
the method described in \cite{Band09,PB09} is isoscattering if the infinite
leads are attached in a way preserving the symmetry of the isospectral
construction \cite{Sawicki10,Sawicki10b}.

Quantum graphs can be considered as idealizations of physical networks in the
limit where the widths of the wires are much smaller than their lengths. They
were successfully applied to model variety of physical problems, see, e.g.,
\cite{Gnutzmann2006} and references cited therein. They can also be realized
experimentally. Recent developments in various epitaxy techniques allowed also
for the fabrication and design of quantum nanowire networks
\cite{Samuelson2004,Heo2008}.

In a seminal work by Hul et al.\ \cite{Hul04} it was shown how quantum graphs
could be successfully simulated by microwave networks. It was demonstrated that
the one-dimensional Schr\"odinger equation for quantum graphs is formally
equivalent to the telegrapher's equation describing microwave networks. For
that reason properties of quantum graphs can be studied experimentally using
microwave networks with the same topology and boundary conditions at the
vertices. Various spectral and scattering properties of microwave networks have
been studied so far \cite{Hul04,Hul2005,Lawniczak2008,Lawniczak2010}.

A quantum graph consists of $n$ vertices connected by $B$ bonds. Each vertex
$i$ of a graph is connected to the other vertices by $v_{i}$ bonds, $v_{i}$ is
called the valency of the vertex $i$. A wave function propagates on each bond
of a graph according to the one-dimensional Schr\"odinger equation. Spectral
properties of a graph are determined by the lengths of bonds connecting
vertices and vertex boundary conditions relating amplitudes of the waves
meeting at each vertex. In the following we consider graphs with two most
physical vertex boundary conditions, the Neumann and Dirichlet ones. The former
impose the continuity and vanishing of the sum of the derivatives calculated at
a vertex $i$ of waves propagating in bonds meeting at $i$. The latter demands
vanishing of the wave function at the vertex.

\begin{figure}[!]
\begin{center}
\rotatebox{0} {\includegraphics[width=0.5\textwidth,
height=0.6\textheight, keepaspectratio]{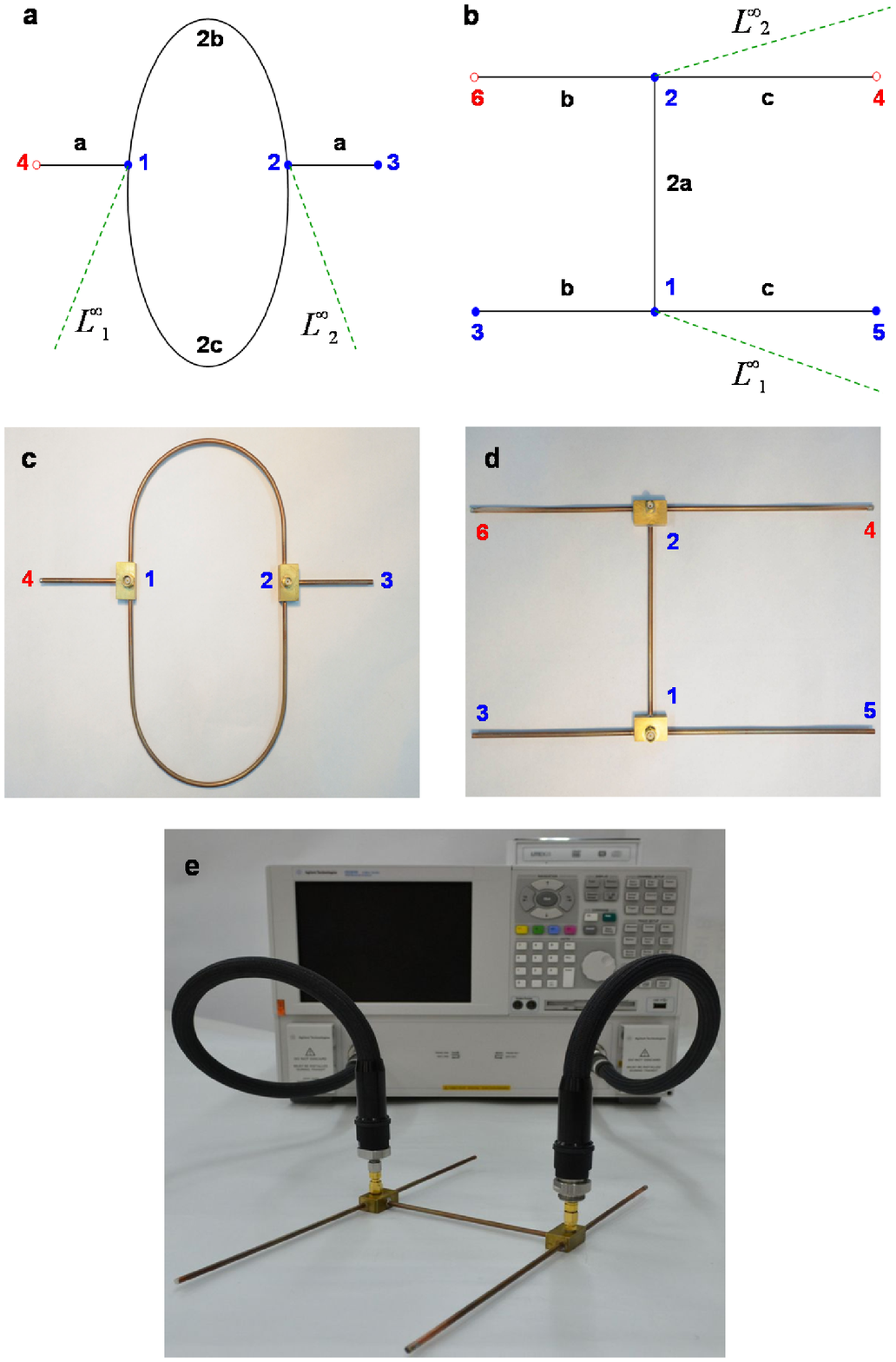}} \caption{A pair of
isoscattering quantum graphs and the pictures of two isoscattering
microwave networks are shown in the panels (a)--(b) and (c)--(d),
respectively. Using the two isospectral graphs, (a) with $n=4$
vertices and (b) with $n=6$ vertices, isoscattering quantum graphs
are formed by attaching the two infinite leads $L^{\infty}_1$ and
$L^{\infty}_2$ (green dashed lines). The vertices with Neumann
boundary conditions are denoted by blue full circles while the
vertices with Dirichlet boundary conditions by red open ones. The
two isoscattering microwave networks with $n=4$ and $n=6$ vertices
which simulate quantum graphs (a) and (b), respectively, are shown
in the panels (c)--(d). The vertices of both networks are
numbered, the numbers are colored (red/blue), that refers to the
(Neumann/Dirichlet) boundary conditions at the vertices. The panel
(e) shows the experimental setup used to measure the two-port
scattering matrix $S(\nu)$ of the networks. In the experiment the
vector network analyzer Agilent E8364B was used. } \label{Fig1}
\end{center}
\end{figure}

In order to test experimentally a negative answer to the modified Mark Kac's
question we consider two graphs shown in Fig.~1a and Fig.~1b. The graphs are
isospectral \cite{Sawicki10}. The isoscattering graphs are obtained from them
by attaching two infinite leads $L^{\infty}_1$ and $L^{\infty}_2$.  Two
corresponding microwave isoscattering networks constructed from microwave
coaxial cables are shown in Figs.~1c and 1d. In order to preserve the same
approximate size of the graphs in Fig.~1a and Fig.~1b and the networks in
Fig.~1c and Fig.~1d, respectively, the lengths of the graphs were rescaled
down to the physical lengths of the networks, which differ from the optical
ones by the factor $\sqrt{\varepsilon}$, where $\varepsilon \simeq 2.08$ is
the dielectric constant of a homogeneous material filling the space between
the inner and the outer leads of the cables.

The graph in Fig.~1a consists of $n=4$ vertices connected by $B=4$ bonds. The
valency of the vertices $1$ and $2$ reads $v_{1,2}=4$ (including leads) while
for the other ones $v_i=1$. At the vertices with numbers $1,2$ and $3$ the
Neumann vertex conditions are satisfied while for the vertex $4$ the
Dirichlet condition is imposed. The second graph (see Fig.~1b) consists of
$n=6$ vertices connected by $B=5$ bonds. At the vertices with numbers $1,2,3$
and $5$ we impose the Neumann vertex conditions, while for the vertices $4$
and $6$ we have the Dirichlet one.

Each system is described in terms of $2\times 2$ scattering matrix
$S(\nu)$:
\begin{equation}
S(\nu)=\left( \begin{array}{cc} S_{1,1}(\nu)&S_{1,2}(\nu)\\
S_{2,1}(\nu)&S_{2,2}(\nu)\end{array} \right) \mbox{,}
\end{equation}
relating the amplitudes of the incoming and outgoing waves of
frequency $\nu$ in both leads.

Since the graphs presented in Fig.~1a and Fig.~1b are isoscattering the
phases of the determinants of their scattering matrices should be equal for
all values of $\nu$:
\begin{equation}\label{2}
\mathrm{Im}\Bigl[\log\Bigl(\det\bigl(S^{(I)}(\nu)\bigr)\Bigr)\Bigr]=
\mathrm{Im}\Bigl[\log\Bigl(\det\bigl(S^{(II)}(\nu)\bigr)\Bigr)\Bigr]\mbox{.}
\end{equation}

In order to measure the two-port scattering matrix $S(\nu)$ we connected the
vector network analyzer (VNA) Agilent E8364B to the vertices $1$ and $2$ of
the microwave networks shown in Fig.~1c and Fig.~1d and performed
measurements in the frequency range $\nu = 0.01-1.7$ GHz. The
connection of the VNA to a microwave network (see Fig.~1e) is equivalent to attaching of
two infinite leads to quantum graphs which means Figs.~1a and 1b correctly
describe the actual experimental arrangement.

The optical lengths of the bonds of the microwave networks had the
following values:

\begin{tabular}{ l r }
 $ a=0.0985\pm 0.0005\mbox{ m, }$ & $2a=0.1970\pm 0.0005\mbox{ m }$ \\
 $ b=0.1847 \pm 0.0005\mbox{ m, }$ & $2b=0.3694 \pm 0.0005\mbox{ m }$ \\
 $ c=0.2420 \pm 0.0005 \mbox{ m, }$ & $2c=0.4840 \pm 0.0005 \mbox{ m }$\\
\label{LConst}
\end{tabular}

At the frequency $\nu = 1.7$ GHz the total optical length of the networks spans
$5.96$ wavelengths of the microwave field.
The uncertainties in the bonds' lengths of the networks are due to the
preparation of Neumann $v_{1,2}=4$ and Dirichlet vertices. In the case of the
first ones the internal leads of the cables were soldered together while the
Dirichlet vertices were prepared by closing the cables with brass caps to
which the internal and external leads of the coaxial cables were soldered.

\begin{figure}[!]
\begin{center}
\rotatebox{0} {\includegraphics[width=0.5\textwidth,
height=0.6\textheight, keepaspectratio]{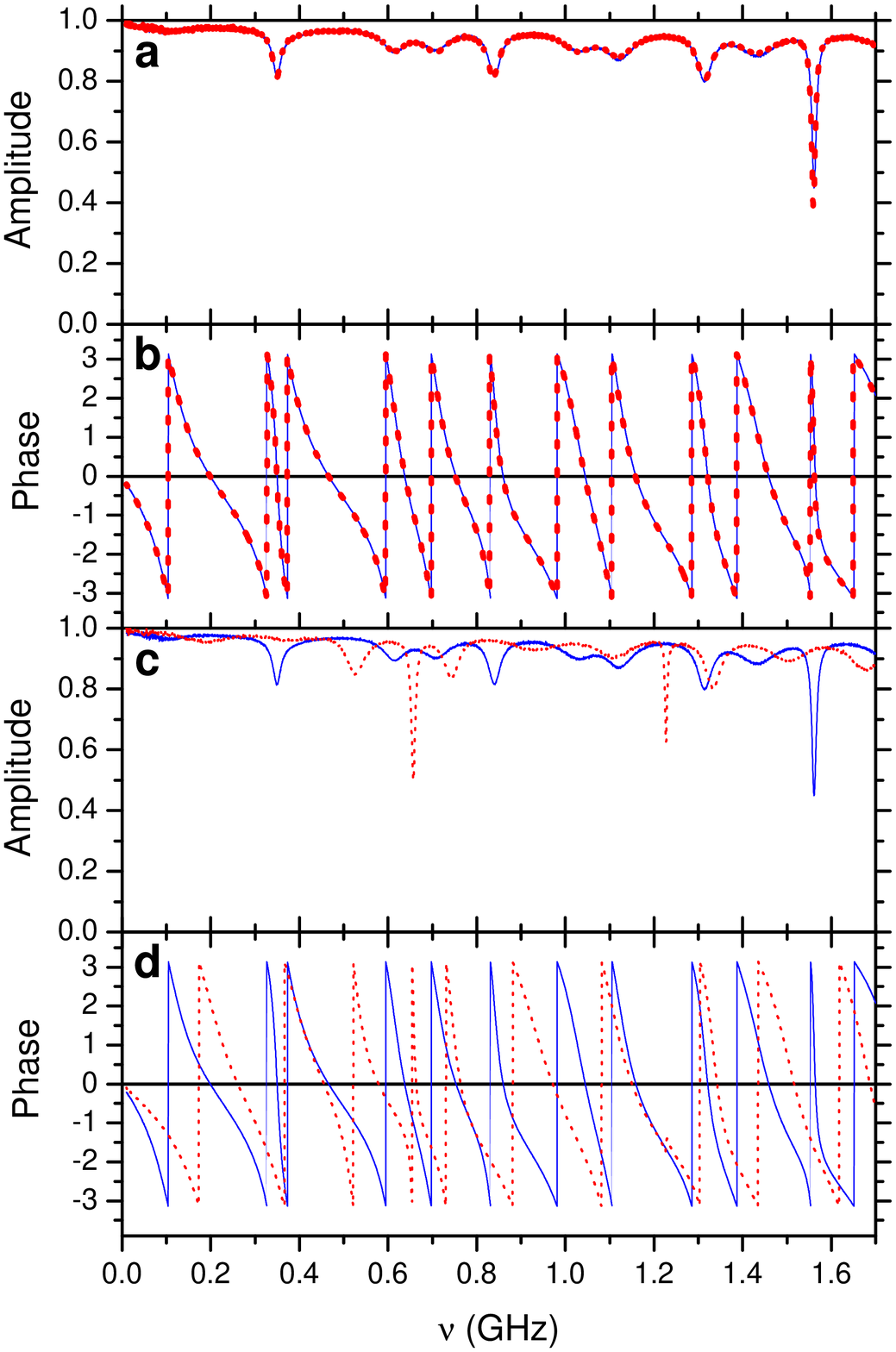}} \caption{(a) The
amplitude of the determinant of the scattering matrix obtained for
the microwave networks with $n=4$ (blue solid line)
 and $n=6$ (red dotted line) vertices. (b) The phase of the
determinant of the scattering matrix obtained for the microwave
networks with $n=4$ (red dotted line) and $n=6$ (blue solid line)
vertices. (c)-(d) The amplitudes and the phases of the
determinants of the scattering matrices, respectively, measured
for the network presented in Fig~1c (blue solid line) and the
modified network Fig~1d (red dashed line), where the Neumann
boundary condition in the vertex $5$ was replaced by the Dirichlet
one. The results are presented in the frequency range $0.01-1.7$
GHz.} \label{Fig2}
\end{center}
\end{figure}

In the case of the  microwave networks, where one deals with losses in the
microwave cables \cite{Hul04}, not only the phase of the determinant $\phi =
\mathrm{Im}\Bigl[\log\Bigl(\det\bigl(S(\nu)\bigr)\Bigr)\Bigr]$ but also the
amplitude $|\det\bigr(S(\nu)\bigl)|$ as well gives an insight into the
resonant structure of the system. The amplitudes and the phases of the
determinants of the scattering matrices of the experimentally studied
networks are shown in Fig.~2a and in Fig~2b, respectively. One sees that
especially for lower frequencies $0.01-1.0$ GHz there is an excellent
agreement between the results obtained for the both networks. The amplitudes
of the determinants are so close to each other that the differences between
them are hardly resolved in Fig.~2a. The phases of the determinants (see
Fig.~2b) are in very good agreement in the full range of the investigated
frequency $\nu= 0.01-1.7$ GHz. In order to demonstrate the sensitivity of the
spectral properties of the networks to the
choice of the boundary conditions we compared the amplitudes and the phases
of the determinants of the scattering matrices measured for the network
presented in Fig~1c (blue solid line) and the modified network Fig~1d (red
dashed line), where the Neumann boundary condition in the vertex $5$ was
replaced by the Dirichlet one. One can easily see in Fig.~2c and Fig.~2d that
such a modification causes a huge departure from the isoscattering
properties.

Our experimental results strongly suggest the impossibility of `hearing' of
the shape of a graph or, in other words, that the question "Are scattering
properties of graphs uniquely connected to their shapes?" has to be answered
in the negative.

Some small differences between the amplitudes appearing for $\nu > 1$ GHz are
due to different lengths of the networks. As it was discussed earlier the
bonds' lengths are known only with a certain accuracy. In order to check the
influence of different bonds' lengths we performed numerical calculations
which took into account also the internal absorption of microwave cables
\cite{Hul04}. We found that at certain realizations of the networks lengths
the results, not shown here, mimic the behavior visible in Fig.~2a.

It was proven by the authors of \cite{Sawicki10} that the graphs considered
in this paper have an additional important property, namely the scattering
matrices of the graphs are conjugated to each other by the following
transplantation relation:

\begin{equation}
\label{Transplantation} S^{(II)}(\nu)=T^{-1}S^{(I)}(\nu)T\mbox{,}
\end{equation}
where $T=\left( \begin{array}{cc}1&-1\\ 1&1\end{array} \right)$. It is worth
noting that the matrix $T$ does not depend on the frequency and the equation
(\ref{Transplantation}) is valid for all values of $\nu$.

\begin{figure}[!]
\begin{center}
\rotatebox{0} {\includegraphics[width=0.5\textwidth,
height=0.6\textheight, keepaspectratio]{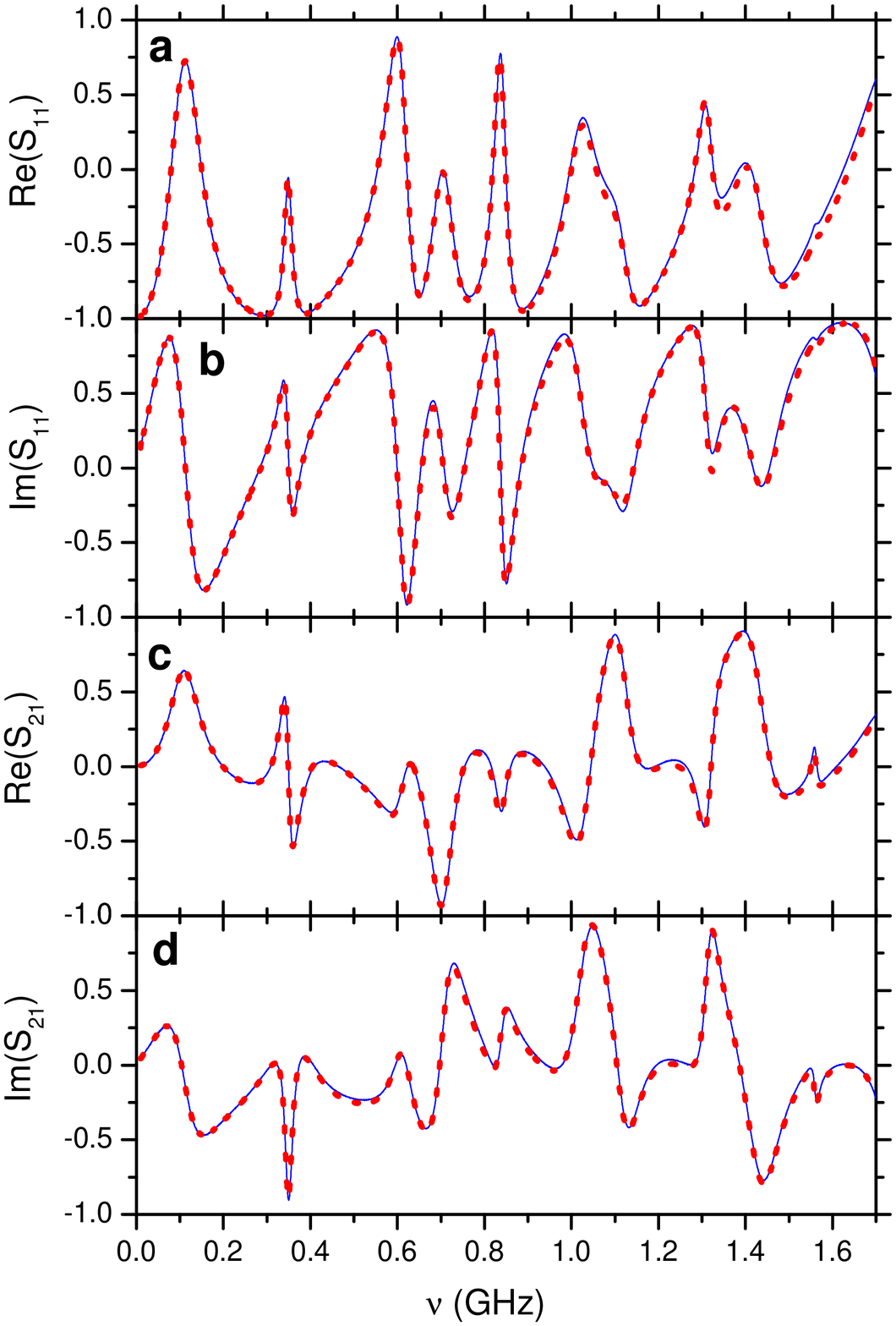}} \caption{(a) The
real and (b) imaginary parts of the matrix element
$\tilde{S}^{(I)}_{1,1}(\nu)$ of the transformed scattering matrix
(blue solid line) for the system with $n=4$ vertices. The obtained
results are compared to the scattering matrix elements
$S^{(II)}_{1,1}(\nu)$ of the graph with $n=6$ vertices (red dotted
line).
 (c) The real and (d) imaginary parts of the matrix element
$\tilde{S}^{(I)}_{2,1}(\nu)$ of the transformed scattering matrix
(blue solid line) for the system with $n=4$ vertices. The results
are compared to the scattering matrix element $S^{(II)}_{21}(\nu)$
of the graph with $n=6$ vertices (red dotted line).
 The results are presented in the frequency range $0.01-1.7$
GHz.} \label{Fig3}
\end{center}
\end{figure}

In order to check the transplantation relation expressed by
equation~(\ref{Transplantation}) we transformed experimentally measured
scattering matrix of the first network
$\tilde{S}^{(I)}(\nu)=T^{-1}S^{(I)}(\nu)T$ and compared it to the scattering
matrix of the second network $S^{(II)}(\nu)$. In Fig.~3 we present the
results for the real and imaginary parts of $S_{1,1}$ and $S_{2,1}$ elements,
respectively. The figure shows clearly that the transplantation relation for
the real and imaginary parts of $S_{1,1}(\nu)$ and $S_{2,1}(\nu)$ elements
works very well. Some small differences seen for $\nu
>1 $ GHz are caused, as previously, by small differences in the cables'
lengths. However, in general, the transformed scattering matrix of the first
network $\tilde{S}^{(I)}(\nu)$ reconstructs very well the scattering matrix
of the second one $S^{(II)}(\nu)$.

The considered microwave networks are obviously dissipative due to the
absorption in the bonds. The loses are proportional to the total length of
bonds, in our case the same for both networks. As it was shown in
\cite{Hul04} loses can be effectively incorporated to the description by
treating the wave number $k$  as a complex
quantity with absorption-dependent imaginary part $\mathrm{Im}\Bigl[k\Bigr]$ and the real part
$\mathrm{Re}\Bigl[k\Bigr] =  2\pi \sqrt{\varepsilon}\nu/ \textsc{c} $,
 where $\textsc{c}$ is the speed of light in vacuum. On the other hand the
authors of \cite{Sawicki10b} proved that the transplantation formula
(\ref{Transplantation}) is satisfied also for complex $k$ (see p. A-152 in
\cite{Sawicki10b}). It was thus reasonable to expect that the influence of
dissipation on the presented results can be neglected and it was indeed the
case. The above theoretical findings were also confirmed in the numerical
calculations (not presented here) which showed that the internal absorption
of the cables does not influence the transplantation relation
(\ref{Transplantation}). Consequently, the validity of the transplantation
relation between the two-port scattering matrices could be experimentally
demonstrated with such a good accuracy as in Fig. 3.

Summarizing, we investigated experimentally scattering properties of two
microwave networks. We showed that the concept of isoscattering graphs was
not only a theoretical idea but it could be also realized experimentally. We
demonstrated that the microwave networks considered in the experiment are
isoscattering, i.e., the phases and amplitudes of the determinant of the
two-port scattering matrices are the same, within the experimental errors,
for all the frequencies considered. In this way we strongly support a
negative answer to the title question about possibility of connecting
uniquely the shapes and scattering properties of graphs. In addition we
checked the validity of the transplantation relation between the two-port
scattering matrices of the two isoscattering microwave networks. It was shown
that this relation allows to reconstruct the scattering matrix of each
investigated network using the scattering matrix of the other one.

Our experimental setup can be successfully used to investigate properties of
any quantum graph, also with highly complicated topology, see, e.g.,
\cite{Lawniczak2008,Lawniczak2010,Hul2011}. Here we showed that they are also
relevant in the study of one of 'abstract' but highly important mathematical
problems of the spectral analysis showing a great research potential of
quantum simulations based on microwave networks.

The authors thank R. Band for critical reading of the manuscript. This work
was supported by the Ministry of Science and Higher Education grant No. N
N202 130239.

\end{document}